\documentstyle[12pt]{article}
\newcommand{\bee}{\begin{equation}}
\newcommand{\ee}{\end{equation}}
\newcommand{\ba}{\begin{array}}
\newcommand{\ea}{\end{array}}
\newcommand{\bea}{\begin{eqnarray}}
\newcommand{\eea}{\end{eqnarray}}

\begin{document}
\thispagestyle{empty}
\begin{flushright}
MPI-Ph/93-87\\
AZPH-TH/93-33\\
November 1993
\end{flushright}
\bigskip\bigskip\begin{center}
{\bf \Huge{Super-Instantons and the Reliability}}\vskip1mm
{\bf \Huge{of Perturbation Theory}}\vskip1mm
{\bf \Huge{in Non-Abelian Models}}
\end{center}
\vskip 1.0truecm
\centerline{\bf
Adrian Patrascioiu}
\vskip5mm
\centerline{Physics Department and Center for the Study of Complex
Systems}
\centerline{University of Arizona, Tucson, AZ 85721, U.S.A.}
\vskip5mm
\centerline{and}
\vskip5mm
\centerline{\bf Erhard Seiler}
\vskip5mm
\centerline{Max-Planck-Institut f\"{u}r
 Physik}
\centerline{ -- Werner-Heisenberg-Institut -- }
\centerline{F\"ohringer Ring 6, 80805 Munich, Germany}
\vskip 2cm
\bigskip \nopagebreak \begin{abstract}
\noindent
In dimension $D\leq 2$ the low temperature behavior of systems enjoying
a continuous symmetry is dominated by super-instantons: classical
configurations of arbitrarily low energy. Perturbation theory in the
background of a super-instanton produces thermodynamic answers for the
invariant Green's functions that differ from the standard ones, but
only in non-Abelian models and only starting at $O(1/\beta^2)$.
This effect modifies the $\beta$-function of the $O(N)$
models and persists in the large $N$ limit of the $O(N)$ models.
\end{abstract}
\vskip 2cm

\newpage\setcounter{page}1


Consider a ${\bf Z}^D$ lattice and let $\Lambda$ be some given finite
subset of it. The non-linear $O(N)$ $\sigma$
model with standard nearest neighbor interaction (s.n.n.i.) at inverse
temperature $\beta$ is defined by the following partition function $Z$:

\bee
  Z=\int\prod_{\langle ij\rangle}\exp(\beta S_i\cdot S_j)\prod_i
\biggl(\delta(S_i^2-1)dS_i\biggr)
\ee
Here S is an $N$-vector, the integration is over the spins inside $\Lambda$
with some boundary conditions (b.c.).

We shall be interested in the behavior of fixed (lattice) distance Green's
functions, such as the expectation value of the energy density
$\langle S(0)\cdot S(1)\rangle$, at large $\beta$. As
$\beta\to\infty$, this finite system will `freeze' into the
one or several configurations which minimize the energy (at given b.c.). For
$\beta$ large but finite, the spins will fluctuate mildly around one of these
classical configurations and thus one can compute the Green's functions
by a standard saddle point expansion. This perturbation theory (PT) approach
produces the correct asymptotic expansion in $1/\beta$; more precisely,
there exists an infinite set of
finite numbers $\{d_k\}$ such that for $\beta$ sufficiently large

\bee
 |\langle S(0)\cdot S(1)\rangle-\sum_{j=0}^k {c_j\over\beta^j}|
< {d_{k+1}\over \beta^{k+1}}
\ee
Here $\{c_k\}$ are the expansion coefficients produced by PT.

Of course both $\{c_k\}$ and $\{d_k\}$
depend upon the linear size $L$ of $\Lambda$ and the
b.c.~used. In particular, they could diverge as $L\to\infty$.
But it could also happen that even though the coefficients $c_k$ have
finite limits, for all $k>k_o$ the left hand side of eq.(2) is not
$o(\beta^{-k})$ uniformly in $L$, i.e.~if multiplied by $\beta^k$ it
is not bounded by a function of $\beta$ alone that vanishes in the limit
$\beta\to\infty$. This would indicate that the $L\to\infty$ limits of the
$c_k$ are not the coefficients of the correct asymptotic expansion
in that limit.
In other words, PT itself cannot be used to establish its correctness in
producing the asymptotic expansion for $\langle S(0)\cdot S(1)\rangle$
at large $\beta$. For that purpose one needs bounds uniform in $L$
on the numbers $\{d_k(L)\}$.
This property of PT was established for the Abelian case $O(2)$ \cite{Bric},
but the authors could not extend the proof to the non-Abelian case
$O(N)$ $N>2$.

The general attitude in the community has been to ignore this issue. In fact
as we tried to emphasize in the past \cite{PaRi,F&F} there are
some good reasons to expect trouble for $D\leq 2$. Firstly in those cases, the
Mermin-Wagner theorem guarantees that the basic assumption
underlying a saddle point approximation, the existence of a dominant
configuration, is violated as $L\to\infty$. Secondly in $1D$, where one can
obtain the true answer using the transfer matrix formalism, one can see
explicitely that for $N>2$, the coefficient of $1/\beta^2$ in the expansion of
$\langle S(0)\cdot S(1)\rangle$ is
finite yet incorrectly given by PT; that means that if one
truncated the series at this order, the remainder would be $O(1/\beta^2)$,
rather than $o(1/\beta^2)$, as required by a putative asymptotic expansion.

In this letter we would like to point out that also in $2D$, the two limits
$\beta\to\infty$ and $L\to\infty$ do not commute for the non-Abelian case
$N>2$. The idea is the following: let $\Lambda$ be a square of size
$L\times L$ and on its
edge freeze the spins along the $x$-axis. Since by the Mermin-Wagner theorem
the infinite volume Gibbs state is $O(N)$ symmetric,
for $L\to\infty$, the expectation value of $\langle S(0)\cdot S(1)\rangle$
should be unchanged if the spin at the origin $S(0)$ is also
frozen in some given direction. So if one believes in PT, one should also
expect that the PT coefficients approach values
independent of whether  $S(0)$ has been frozen or not. As it will be seen,
this is precisely what happens for $O(2)$ but not for $O(N)$ $N>2$ starting at
$O(1/\beta^2)$.

First let us discuss the massless Gaussian, described by the partition
function

\bee
  Z=\int\prod_{\langle ij\rangle}\exp(-{1\over 2}
(\varphi_i-\varphi_j)^2)\prod_i d\varphi_i
\ee
It too possesses a continuous symmetry, which in $2D$ cannot be broken
spontaneously. Consequently, if we fix $\varphi=0$ at the edge of
the square and
measure an invariant observable, such as
$D(0,1)=(\varphi(0)-\varphi(1))^2$, for $L\to\infty$ the result should become
independent of whether $\varphi(0)$ is fixed or not.
In particular even if $\varphi(0)=0$, $D(0,1)$ should converge to
$1/2$ for $L\to\infty$.

This is clearly the case, as it can be seen from the data in Tab.1 (line 1).
These data were produced by a Monte Carlo study of this model using the
over-relaxation procedure. Besides $D(0,1)$ we measured certain Green's
functions
of the massless Gaussian with these b.c., which will be needed to compute
the PT expansion of  $\langle S(0)\cdot S(1)\rangle$
in $O(N)$ - see below. We increased the number
of measurements with $L$; as an example, for $L=80$ we took $100,000$
measurements,
separated by 5 sweeps of the lattice. Similarly, the number of thermalization
sweeps increased with $L$, reaching $10,000$ for $L=80$. We estimated the
error by a standard binning procedure.

Let us return now to the case of interest of the
$O(N)$ models. Firstly, although
it is obvious, we verified that as we increase $L$,
 $\langle S(0)\cdot S(1)\rangle$ does become
independent of what we do with $S(0)$. We studied both $O(2)$
and $O(3)$. The spins at the edge of the square were fixed along the $x$-axis.
The spin at the origin
$S(0)$ was either free (`Dirichlet b.c.') or also frozen along the $x$-axis
(`super-instanton b.c.'). We performed a Monte Carlo study using a cluster
algorithm. Some typical results are shown in Tab.2, where we give also the
value of  $\langle S(0)\cdot S(1)\rangle$
at the same $\beta$ using periodic b.c. The results indicate
that $\langle S(0)\cdot S(1)\rangle$
does become independent of the b.c. as $L\to\infty$, although
the approach is slower for the super-instanton b.c..

Next let us see what PT would predict for super-instanton b.c. Using the
standard parametrization

\bee
          S_i=(\pi_i,\sqrt{1-\pi_i^2}) ,
\ee
then expanding in $\pi$, one obtains

\bea
 \langle S(0)\cdot S(1)\rangle&=&1-{(N-1)\over 2\beta} G(1,1) \nonumber\\
 &-&{1\over\beta^2}(N-1)^2\biggl[{(G(1,1)^2\over 8}
-{t\over 4}\biggr] \nonumber\\
        &+&(N-1)\biggl[-{G(1,1)^2\over 4}+{v\over 2}\biggr]+...
\eea
where

\bea
 G(i,j)&=&\langle\pi(i)\cdot\pi(j)\rangle, \nonumber\\
      t&=&\sum_{<i,j>}\bigl[G(1,i)^2-G(1,j)^2\bigr]
             \bigl[G(i,i)-G(j,j)\bigr], \nonumber\\
      v&=&\sum_{<i,j>}\bigl[G(1,i)-G(1,j)\bigr]
       \biggl\{G(1,j)\bigl[G(i,j)-G(j,j)\bigr] \nonumber\\
       &-&G(1,i)[G(i,j)-G(i,i)]\biggr\}
\eea
The super-instanton b.c. on $S$
($\pi=0$ along the edge and at 0), imply that $G(i,j)$
in eq.(6) is precisely the Green's function of the massless Gaussian studied
above. From that study we already know that the term of $O(1/\beta)$
will converge
to a value independent of the b.c. as $L\to\infty$. To study the term of
$O(1/\beta^2)$ we computed $t$ and $v$ numerically and report
the values obtained in Tab.1.

To interprete these data, let us recall that if instead of the super-instanton
b.c. one would use periodic b.c., one would obtain \cite{Has}

\bee
        \langle S(0)\cdot S(1)\rangle
=1-{N-1\over\beta} D(1)-{N-1\over 2\beta^2}D(1)^2+...
\ee
Here $D(1)={1\over 2D}(1-{1\over L^2})$ and we have neglected a term of
$O(1/\beta^2)$ coming
from the elimination of the zero mode which vanishes very rapidly with
increasing $L$. Since eq.(5) contains a term proportional to $(N-1)^2$, absent
in eq.(7), the two would agree only if $t$ converged to $1/8$. In fact our
data presented in Tab.1 suggest that $t$ converges to $0.035$ or less.
On the contrary, agreement of eqs.(5) and (7) for $N=2$ requires that
$v-t/2$ converges to $1/8$ and our
data certainly seem consistent with that.

The fact that $v-t/2$ goes to $1/8$ can also be proven analytically.
It follows after some algebraic manipulations that

\bee
v-t/2=\sum_{<i,j>}\biggl(-G(i,j)+{1\over 2}G(i,i)+{1\over 2}G(j,j)
\biggr)\biggl(G(0,i)-G(1,i)-G(0,j)+G(1,j)\biggr)
\ee
It can be justified to take the limit $L\to\infty$ under the sum, and
then it is seen easily by a summation by parts that it becomes $1/8$.
On the other hand, the same procedure is not justified for the sums
defining $v$ and $t$ separately (because the sums are not convergent uniformly
in $L$); one would obtain $t=1/8$ and $v=3/16$, which would agree
with Hasenfratz's formula (7), but clearly disagrees with our data.

Therefore our results suggest that if in $2D$ one takes the limit $L\to\infty$
termwise in the PT expansion, the result depends upon the b.c.~used, but only
for non-Abelian models and then only starting at $O(1/\beta^2)$. Since the
correct asymptotic expansion has to be independent of b.c. by the
Mermin-Wagner theorem, one must conclude that taking
$L\to\infty$ termwise is not a legitimate procedure. Of course it could happen
that in $2D$ this procedure gives the correct infinite volume answer for
certain b.c., just as it does in $1D$ if one uses free b.c.. For instance it
has been argued that periodic b.c. must be giving the correct infinite volume
answer because they agree with the $1/N$ expansion. Unfortunately, as we
explained before \cite{F&F}, this argument is not correct because it
involves again an a priori illegitimate interchange of two limits: $\beta\to
\infty$ and $N\to\infty$. So while it may happen that in $2D$ periodic
b.c. do give
the correct infinite volume PT coefficients, to our knowledge there is no
support for that belief at the present time.

We also computed the PT coefficients to the 2-point function at distances
$x$ larger than one lattice unit, and found that with super-instanton b.c.~one
obtains a different answer in the thermodynamic limit for $N>2$. Since
for large $x$ ($x>4$) this modification grows with $x$ like
$G(x,x)^2$, it will
give rise to a $\beta$-function whose leading coefficient is different
from the one found in the literature (obtained by using PT with periodic b.c.).
This fact casts doubt upon the alleged asymptotic freedom of the $O(N)$
$N>2$ models. Details will be reported elsewhere.

Over the years it has been argued that in theories which are
perturbatively asymptotically free,
PT should be applicable at short distances, where the system is well
ordered. In fact starting at $O(1/\beta^2)$, as it can be seen from eqs.(5)
and (6), a PT computation involves sums over all distances. Consequently our
finding that the trouble starts only from $O(1/\beta^2)$
is quite understandable.
The trouble does not arise in $O(2)$ because there, by parametrizing the spin
as $(\cos(\varphi),\sin(\varphi))$, the
Gibbs measure becomes a function of $\nabla\varphi$ only.

Before concluding let us explain why we called these b.c.~super-instanton b.c..
Let $\Lambda$ be (the closest lattice domain to) a disk of radius R.
Let us use the following spin parametrization

\bee
    S_i=(\pi_i,\sqrt{1-\pi_i^2}\sin(\varphi_i),
\sqrt{1-\pi_i^2}\cos(\varphi_i))
\ee
where the $\pi_i$ are now $(N-2)$-vectors.
We choose as b.c. $\pi=0$ along the edge and at $0$, while
$\varphi(0)=0$ but $\varphi=c>0$
along the edge. To do PT we need first the configuration of minimal energy
with these b.c.: it is some lattice version of the following continuum
expression

\bea
   \varphi(r,\theta)&=&c\log r/\log R \nonumber\\
           \pi(r,\theta)&=&0
\eea
Please notice that this classical configuration has a vanishing energy as
$R\to\infty$, hence the name super-instanton.

More precisely, twenty years ago Kosterlitz and Thouless (KT) \cite{KT}
argued that at large $\beta$ the partition function is dominated by the
classical configurations, which have since been called instantons.
According to KT in $O(2)$ the dominant configurations
are vortices and their energy is actually $O(\log(R))$
(hence they are bound), while in $O(3)$ they are `hedgehogs' of energy
$O(R^0)$, which create exponential decay no matter how large $\beta$ is.
In fact, as we see now, the super-instantons are the dominant configurations
in all $O(N)$ models, since their energy vanishes for $R\to\infty$. They
occur for all $D\leq 2$ and are the `enforcers' of the Mermin-Wagner theorem
(a related idea underlies Pfister's \cite{Pfister} proof of that theorem):
Let us for simplicity work in the continuum, with
$a$ playing the role of an UV cutoff, and use the fact that the configuration
of minimal energy has a $\varphi$ that only depends on the radius $r$. Then

\bea
    c&=&\int_a^R dr {d\over dr}\varphi \nonumber \\
    E&=&2\pi\int_a^R dr r^{D-1}({d\over dr}\varphi)^2  .
\eea
With the scalar product

\bee
             (f,g)= \int_a^R drr^{D-1}f^*g
\ee
we obtain

\bee
              c=(r^{1-D},{d\over dr}\phi)
\ee
and by Schwarz's inequality

\bee
    c\leq ||r^{1-D}||\sqrt E=
\sqrt{{R^{2-D}-a^{2-D}\over 2-D}}\sqrt{E}
\ee
This gives a lower bound for $E$ that goes to $0$ as $R\to\infty$,
whereas for $D>2$ it remains bounded away from $0$.
Therefore one can have super-instantons ($E\to 0$ yet $c\geq c_o>0$)
only for $D\leq 2$.

The arguments above suggest that at low temperature, the typical configuration
in the $O(N)$ model in dimension $D\leq 2$ is a gas (or liquid)
of super-instantons. The remarkable fact is that
we had reached the same conclusion two years ago, when we investigated these
models by mapping them into a certain percolation problem \cite{Pat,Lat92};
namely if we consider a certain patch of the sphere $S^{N-1}$
of angular opening $O(1/\sqrt\beta)$ and ask what is its inverse
image in a typical
configuration, our percolation arguments indicated that the answer must be
that it forms rings of arbitrarily large size (more precisely, neither the
inverse image of the patch nor of its complement percolates).
This super-instanton like structure was found to be
responsible for the absence of exponential decay in all
$O(N)$ models at $\beta$ sufficiently large. Now we see that it may also be
responsible for the failure of PT in non-Abelian models. Indeed imagine doing
PT in the background of a super-instanton. Even though the new couplings
induced by this external field vanish as $1/(\log R)^2$, the potential IR
divergences present in PT could produce nonvanishing effects as $R\to\infty$.
In fact the effect could happen only for $N>2$ ($N=2$ involves only
$\nabla\phi$,
which is IR finite) and only starting at $O(1/\beta^2)$ since
we must compensate
two powers of $1/\log R$ - this is in agreement
with the findings reported above,
which represented PT in the background of the trivial super-instanton.

A.P.~would like to acknowledge the hospitality of the IPN of the
Universit\'e de Paris Sud and of the Werner-Heisenberg-Institute.

\vfill\eject

\noindent
{\bf Tab.1:} {\it The PT coefficients to order} $1/\beta^2$ {\it of}
$\langle S(0)\cdot S(1)\rangle$ {\it in the s.n.n.i.} $O(N)$
{\it models with super-instanton b.c.

  \medskip
  \vbox{\offinterlineskip\halign{
  \strut\vrule#&\quad $#$\quad&\vrule\hskip1pt\vrule#&&\quad $#$\hskip5pt
  &\vrule#\cr
   \noalign{\hrule}
   &L&&  10&&  20&&  40&&  60&&  80&\cr
   \noalign{\hrule\vskip1pt\hrule}
   &G(1,1)&&.3752(10)&&.4006(12)&&.4143(8)&&.4211(13)&&.4314(22)&\cr
   \noalign{\hrule}
   &8v-4t&&.657(32)&&.709(40)&&.772(18)&&.792(18)&&.833(74)&\cr
   \noalign{\hrule}
   &8t&&.345(46)&&.296(47)&&.305(18)&&.284(26)&&.317(77)&\cr
   \noalign{\hrule}}}

%

\vskip2cm
\noindent
{\bf Tab.2a:} $\langle S(0)\cdot S(1)\rangle$ {\it for the s.n.n.i.}
$O(2)$ {\it model with periodic (per.), Dirichlet
(Dir.) and super-instanton (s.i.) b.c. (Monte Carlo data taken at
$\beta=1.1$; the data point for $L=256$ is taken from R.Gupta and
C.Baillie,} {\sl Phys.Rev.} {\bf B45} (1992) 2883.)

  \medskip
  \vbox{\offinterlineskip\halign{
  \strut\vrule#&\quad $#$\quad&\vrule\hskip1pt\vrule#&&\quad $#$\hskip5pt
  &\vrule#\cr
   \noalign{\hrule}
   &L&&  20&&  40&&  80&&  256&\cr
   \noalign{\hrule\vskip1pt\hrule}
   &per.&&.7177(6)&&.7155(3)&&.7146(2)&&.7138(1)&\cr
   \noalign{\hrule}
   &Dir.&&.717(2)&&.712(2)&&     &&      &\cr
   \noalign{\hrule}
   &s.i.&&.774(3)&&.768(3)&&.755(4)&&     &\cr
   \noalign{\hrule}}}
   \vskip5mm

\noindent
{\bf Tab.2b:} $\langle S(0)\cdot S(1)\rangle$ {\it for the s.n.n.i.}
$O(3)$ {\it model with periodic (per.), Dirichlet
(Dir.) and super-instanton (s.i.) b.c. (Monte Carlo data taken at
$\beta=2.0$; the data point for $L=1024$ is taken from J.Apostolakis et al,}
{\sl Phys.Rev.} {\bf D43} (1991) 2687.)

  \medskip
  \vbox{\offinterlineskip\halign{
  \strut\vrule#&\quad $#$\quad&\vrule\hskip1pt\vrule#&&\quad $#$\hskip5pt
  &\vrule#\cr
   \noalign{\hrule}
   &L&&  20&&  40&&  80&&  1024&\cr
   \noalign{\hrule\vskip1pt\hrule}
   &per.&&.7272(5)&&.7257(3)&&.7253(2)&&.72511(3)&\cr
   \noalign{\hrule}
   &Dir.&&.728(2)&&.726(2)&&     &&      &\cr
   \noalign{\hrule}
   &s.i.&&.776(3)&&.766(3)&&.753(4)&&      &\cr
   \noalign{\hrule}}}
   \vskip2cm


\begin{thebibliography}{99}
%
\bibitem{Bric}{J.Bricmont, J.-R.Fontaine, J.L.Lebowitz, E.H.Lieb and
T.Spencer, {\sl Commun.Math.Phys.} {\bf 78} (1981) 545.}
Academic Press, London - New York 1972.
%
\bibitem{PaRi} A.Patrasciou and J.-L. Richard, {\sl Phys. Lett.} {\bf 149B}
(1984) 167.
%
\bibitem{F&F}A.Patrascioiu and E.Seiler, {\it The Difference between Abelian
and Non-Abelian Models: Fact and Fancy}, preprint MPI-Ph/91-88.
%
\bibitem{Has} {P.Hasenfratz, {\sl Phys.Lett.}{\bf B141} (1984)385.}
New York etc. 1989.
%
\bibitem{KT} {J.M.Kosterlitz and D.J.Thouless, {\sl J. Phys. (Paris)}
{\bf 32} (1975) 581.}
%
\bibitem{Pfister} C.Pfister, {\sl Commun.Math.Phys.} {\bf 79}
(1981) 181.
%
\bibitem{Pat}A.Patrascioiu, {\it Existence of Algebraic Decay in
 non-Abelian Ferromagnets}, University of Arizona preprint AZPH-TH/91-49.
%
\bibitem{Lat92}A.Patrascioiu and E.Seiler, {\it Percolation Theory and
the Existence of a Soft Phase in 2D Spin Models}, {\sl Nucl.Phys.B.(Proc.
Suppl.)} {\bf 30} (1993) 184.
%
\end{thebibliography}
\end{document}